\documentclass[aps,prd,floatfix,superscriptaddress,preprintnumbers,showpacs,showkeys]{revtex4}
\usepackage{amssymb}
\usepackage{amsmath}
\usepackage{amsfonts}
\usepackage{epsfig}
\usepackage{graphicx}
\usepackage{tabularx}
\newcommand{\seq}{\begin{subequations}}
\newcommand{\sen}{\end{subequations}}
\newcommand{\eq}{\begin{eqnarray}}
\newcommand{\en}{\end{eqnarray}}

\def\shiftdown#1{#1\llap{\lower.04ex\hbox{#1}}}

\begin{document}

\title{Role of QCD compositeness in the production of scalar and tensor mesons \\ 
through single-photon annihilation $e^+ e^- \to \gamma^* \to \gamma S(T)$}

\author{Alexandr G. Chumakov} 
\affiliation{Laboratory of Particle Physics,
Tomsk Polytechnic University, 634050 Tomsk, Russia} 
\author{Thomas Gutsche}
\affiliation{Institut f\"ur Theoretische Physik, Universit\"at T\"ubingen,
Kepler Center for Astro and Particle Physics,\\
Auf der Morgenstelle 14, D-72076 T\"ubingen, Germany}
\author{Valery E. Lyubovitskij} 
\affiliation{Institut f\"ur Theoretische Physik, Universit\"at T\"ubingen,
Kepler Center for Astro and Particle Physics,\\
Auf der Morgenstelle 14, D-72076 T\"ubingen, Germany}
\affiliation{Departamento de F\'\i sica y Centro Cient\'\i fico
Tecnol\'ogico de Valpara\'\i so-CCTVal, Universidad T\'ecnica
Federico Santa Mar\'\i a, Casilla 110-V, Valpara\'\i so, Chile}
\affiliation{Department of Physics, Tomsk State University,
634050 Tomsk, Russia}
\affiliation{Laboratory of Particle Physics,
Tomsk Polytechnic University, 634050 Tomsk, Russia}
\author{Ivan Schmidt}
\affiliation{Departamento de F\'\i sica y Centro Cient\'\i fico
Tecnol\'ogico de Valpara\'\i so-CCTVal, Universidad T\'ecnica
Federico Santa Mar\'\i a, Casilla 110-V, Valpara\'\i so, Chile}

\date{\today}

\begin{abstract}

We study the exclusive production of scalar $S = 0^{++}$ and 
tensor $T = 2^{++}$ mesons through single-photon annihilation 
$e^+ e^- \to \gamma^* \to \gamma S(T)$. 
Using QCD compositeness of the involved hadrons considered as quark-antiquark 
systems, the prediction for the scaling 
of the differential cross sections  of these processes is
$d\sigma/dt \sim 1/s^3$ at large $s$. 
We further derive the scaling of the $\gamma^\ast \to \gamma S$ and 
$\gamma^\ast \to \gamma T$ transition form factors: 
$F_{\gamma^\ast\gamma S}(s) \sim 1/s$ and 
$F_{\gamma^\ast\gamma T}(s) \sim 1/s^2$. 
Results for the respective cross sections of the scalar and tensor meson production
are presented. Note, when scalar and tensor mesons are considered 
as tetraquark systems of two tightly bound color diquarks,  
corresponding to them transition form factors and differential cross sections 
have the same falloffs as in case of quark-antiquark picture. 
For other tetraquark or two-hadron molecules configurations the transition 
form factors $F_{\gamma^\ast\gamma S(T)}(s)$ and the differential cross section 
$d\sigma/dt$ have additional $1/s$ and $1/s^2$ falloffs, respectively. 

\end{abstract}



\maketitle

\section{Introduction} 

In this paper we present a study of the exclusive 
production of scalar $S = 0^{++}$ and tensor $T = 2^{++}$ mesons through 
single-photon annihilation $e^+ e^- \to \gamma^* \to \gamma S(T)$, which 
is described by the diagram in Fig.~1. Here $p_1$, $p_2$, $q$, $q_1$, and 
$q_2$ are the momenta of the initial electron, positron, intermediate photon, 
final photon, and scalar (tensor) meson, respectively. 

\begin{figure}
\begin{center}
\epsfig{figure=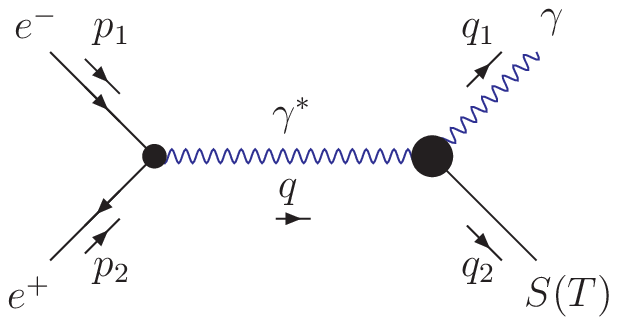,scale=1}
\noindent
\caption{The single-photon annihilation $e^+ e^- \to \gamma^* \to \gamma S(T)$. 
\label{fig:epem_ST}}
\end{center}
\end{figure}
 
The main idea of the paper is to make predictions for the integral 
cross section of the production of scalar and tensor mesons 
in the reaction $e^+ e^- \to \gamma^* \to \gamma S(T)$ in whole region of 
variable $s = q^2$ (i.e., without restriction to small or high values of s). 
Our strategy is the following. Using the QCD prediction for power scaling 
of differential cross section of $e^+ e^- \to \gamma^* \to \gamma S(T)$ 
we constrain the power scaling of the $\gamma^* \to \gamma S(T)$ 
transition form factors (denoted by big black vertex in Fig.1). 
Explicit form of the form factors can be finally 
fixed using available data or available results from phenomenological approaches. 
Finally, we make the numerical analysis of the integral cross sections 
of $e^+ e^- \to \gamma^* \to \gamma S(T)$ processes. Our predictions are valid 
in whole region of $s$ and will be useful for planning experiments at 
electron-positron colliders. 

The scaling results for the exclusive cross section 
$e^+ e^- \to \gamma^* \to \gamma S(T)$ must be consistent 
with the leading-twist quark fixed-angle counting 
rules~\cite{Brodsky:1973kr}-\cite{Lepage:1980fj}:
$\frac{d\sigma}{dt}(A+ B \to C +D) \propto F(\theta_{CM})/
s^{N-2}$, where $N= N_A + N_B + N_C + N_D $ is the total twist or
number of elementary constituents and $F(\theta_{CM})$ is the square of 
the fixed $\theta_{CM}$--angle amplitude. 
In our case when scalar and tensor mesons are considered as $q\bar q$ systems 
we have $N - 2 = 3$ resulting in the 
scaling behavior $\frac{d\sigma}{dt} \sim 1/s^3$. Alternatively, the scalar 
and tensor mesons could have tetraquark or two-hadron molecule structure. 
Application of the QCD compositeness for the tetraquarks 
has been done in Refs.~\cite{Blitz:2015nra,Brodsky:2015wza,Brodsky:2016uln}.  
It was shown that when tetraquarks are systems of two tightly bound color 
diquarks, the differential cross section has the same falloff 
as in the case of quark-antiquark picture. For other tetraquark 
configurations and hadronic molecules the differential cross section $d\sigma/dt$ 
have additional $1/s^2$ falloff for each exotic state in the final state. 

Note that recently, in Ref.~\cite{Brodsky:2016uln},             
QCD compositeness was successfully applied to the study of production of 
vector mesons --- of single and double 
vector meson production in $e^+ e^-$ annihilation. 
It was shown that both the differential and integral cross sections scale
as $1/s$. The reason for this behavior is that the corresponding amplitudes 
are dominated by the spin$-\frac{1}{2}$ electron exchange in the $t$ and $u$ 
channels, which gives an extra factor of $s^{\frac{1}{2}}$. 
The results for the production of vector mesons have been generalized 
in Ref.~\cite{Brodsky:2016uln}  
to the exclusive double-electroweak vector-boson 
annihilation processes, accompanied by the forward production of hadrons, 
such as $e^+ e^- \to Z^0 V^0$ and $e^+ e^- \to W^-\rho^+$, and 
the exclusive production of exotic hadrons --- 
tetraquarks. It motivates to continue study of production of other 
meson states using principle of the QCD compositeness. In particular, 
in this paper we focus on production of scalar and tensor mesons. 

In the case of $\gamma S(T)$ production the dominant diagram is the 
$s$ channel process $e^+ e^- \to \gamma^* \to \gamma S(T)$.
Such processes have been searched for experimentally by the SND Collaboration 
at the VEPP-2M $e^+e^-$ collider in Ref.~\cite{Achasov:2011zza}, 
and studied in Refs.~\cite{Achasov:2013daa,Achasov:2015pba}. 
In particular, in Refs.~\cite{Achasov:2013daa,Achasov:2015pba} 
the production of scalar mesons has been studied using the vector 
dominance model (VDM) and the kaon loop model, while in the case of 
tensor mesons only the VDM 
was used in the analysis. It was found that the kaon loop model gives a good 
description of data up to $\sqrt{s} \simeq 2$ GeV. 

We stress 
that the claim presented in Refs.~\cite{Achasov:2013daa,Achasov:2015pba} 
which states that the integral cross section for $\gamma S$ and $\gamma T$ 
productions should scale as $1/s$ does not appear to be correct since 
it contradicts the results of QCD compositeness (quark counting rules). 
The integrated cross section should scale as $1/s^2$, 
while the differential cross 
section should scale as $1/s^3$. Our observation is consistent with 
Ref.~\cite{Brodsky:2016uln}, where the falloff of 
the differential cross section for the production of photon and vector 
meson is $1/(s t^2)$, i.e., a generic Mandelstam variable with power $-3$. 
The difference of $\gamma V$ and $\gamma S$ production is clear: 
in the first case the falloff of the differential cross section 
is $1/(s t^2)$, while in the second case it is $1/s^3$. 

As we mentioned above, 
using the specific falloff of the differential cross section for 
$\gamma S(T)$ production we can constrain the falloff of 
the $\gamma^* \to \gamma S$ and $\gamma^* \to \gamma T$ 
transition form factors at large $s$ as $F_{\gamma^*\gamma S}(s) \sim 1/s$ 
and $F_{\gamma^*\gamma T}(s) \sim 1/s^2$, respectively. 
Note that the $1/s$ and $1/s^2$ scalings of the $F_{\gamma^*\gamma S}(s)$ 
and $F_{\gamma^*\gamma T}(s)$ form factors  
are to be expected, because in the Euclidean region 
the $\gamma^*(Q^2) \gamma \to f_0(980)$ and 
$\gamma^*(Q^2) \gamma \to f_2(1270)$, $a_2(1320)$ 
transition form factors have similar power scaling, 
$1/Q^2$~\cite{Kroll:2016mbt} and $1/Q^4$~\cite{Achasov:2015pha} 
at large $Q^2$, respectively. 

Note that when the scalar mesons $f_0(980)$ and $a_0(980)$ 
are considered as tetraquarks, there are two 
scenarios for the quark configuration --- hadronic molecular (HM)
and color diquark-antidiquark (CD) configuration. 
In Refs.~\cite{Blitz:2015nra,Brodsky:2015wza}
it was shown that for the a system of two tightly bound diquark 
in the CD scenario the corresponding form factors, involving these states,  
have the same falloff as form factors involving quark-antiquark systems, 
otherwise each tetraquark state costs an additional $1/s$ falloff. 
This model-independent feature was recently confirmed 
in Ref.~\cite{Brodsky:2016uln}, 
where electron-positron annihilation into single and double tetraquarks 
was considered in a soft-wall AdS/QCD approach. 
It means that in case of the $f_0(980)$ and $a_0(980)$ states 
two possible scenarios for their inner structure - quark-antiquark 
configuration and tetraquark system with two tightly bound color diquarks 
we have $1/s$ scaling of the corresponding $\gamma^* \to \gamma S$ 
transition form factor at large $s$, which is consistent with results 
of Ref.~\cite{Kroll:2016mbt}. Therefore, in this paper we consider 
these scenarios for the $f_0(980)$ and $a_0(980)$ states. 

In the following we proceed to set up the effective formalism with the aim 
of describing the $e^+ e^- \to \gamma^* \to \gamma S(T)$ annihilation 
reactions constrained by the scaling behavior for large $s$. 
We introduce form factors in these transitions which are minimally 
parametrized to reproduce the available cross section data and results 
of other phenomenological approaches.  
We also give predictions for $e^+ e^-$ annihilation reactions of a variety
of scalar and tensor mesons. 

\section{Formalism and Results} 

Our starting point is the effective Lagrangian
for the couplings of scalar ($S$) and tensor $(T)$ mesons to two photons.  
In Table 1 we specify quantum numbers, masses and 
their two-photon decay widths, which will be used in numerical analysis. 
The off-shell behavior of one photon 
annihilating into $\gamma S$ and $\gamma T$ is parametrized 
by the form factors $F_{\gamma^*\gamma S}$ and $F_{\gamma^*\gamma T}$ 
included in the Lagrangians: 
\eq
{\cal L}_{\gamma^*\gamma S}(x) &=&
\frac{e^2}{4} \, g_{S\gamma\gamma} \, S(x)  \, F_{\mu\nu}(x) \,
\int d^4y \, {\cal F}_{\gamma^*\gamma S}(x-y) F^{\mu\nu}(y)\,, \nonumber\\
{\cal L}_{\gamma^*\gamma T}(x) &=& \frac{e^2}{2} \, g_{T\gamma\gamma}  \,
T_{\mu\nu}(x)  \, F^\mu_{\ \alpha}(x) \,
\int d^4y \, {\cal F}_{\gamma^*\gamma T}(x-y) \, F^{\alpha\nu}(y)\,,
\nonumber\\
{\cal F}_{\gamma^*\gamma S(T)}(x-y)&=& 2 F_{\gamma^*\gamma S(T)}(x-y) - 
\delta^4(x-y)\,,  
\en
$F_{\mu\nu} = \partial_\mu A_\nu - \partial_\nu A_\mu$ is the stress 
tensor of the electromagnetic field, $S$ is the $J^{PC} = 0^{++}$ scalar 
field, 
and $T_{\mu\nu}$ is the $J^{PC} = 2^{++}$ tensor field, which is a symmetric 
($T_{\mu\nu} = T_{\nu\mu}$), traceless ($T_{\ \mu}^\mu = 0$) rank-2 tensor 
obeying the transversity condition 
($\partial^\mu T_{\mu\nu} = 0$)~\cite{Giacosa:2005zt}. 
Here $e$ is the elementary electric charge. The on-shell couplings 
$g_{S\gamma\gamma}$ and $g_{T\gamma\gamma}$
define the two-photon decay widths 
of the $S$ and $T$ mesons: 
\eq
\Gamma(S \to \gamma\gamma) = \frac{\pi}{4} \, \alpha^2 \,
g_{S\gamma\gamma}^2 \, M_S^3 \,, \quad
\Gamma(T \to \gamma\gamma) = \frac{\pi}{20} \, \alpha^2 \,
g_{T\gamma\gamma}^2 \, M_T^3 \,,
\en
where $M_S$ and $M_T$ are the respective meson masses.  
The relativistic form factors which take into account the off-shellness 
of the virtual $\gamma^*$ photon are normalized to 1 at $s=0$. Note that 
the idea to introduce a form factor for the $\gamma^* \to \gamma T$ 
transition was originally proposed in Ref.~\cite{Achasov:2015pba}.  

\begin{table}[ht]
\begin{center}
\caption{Scalar and tensor mesons.} 
\label{tab:bb}
\def\arraystretch{1.25}
\begin{tabular}{|c|c|l|l|}
\hline
Name & $I^G (J^{PC})$ & Mass (MeV) & $\Gamma(H \to \gamma\gamma)$ (keV) \\
\hline\hline
$f_0(980)$   & $0^+ (0^{++})$ & $990 \pm 20$ & $0.31^{+0.05}_{-0.04}$ \\
$f_0(1370)$  & $0^+ (0^{++})$ & $1370^{+130}_{-170}$ & \\
$f_0(1500)$  & $0^+ (0^{++})$ & $1504 \pm 6$ & \\
$f_0(1710)$  & $0^+ (0^{++})$ & $1723^{+6}_{-5}$& \\
$a_0(980)$   & $1^- (0^{++})$ & $980 \pm 20$ & $0.30 \pm 0.10$        \\
$a_0(1450)$  & $1^- (0^{++})$ & $1474 \pm 19$         & \\
$a_0(1950)$  & $1^- (0^{++})$ & $1931 \pm 14 \pm 22$  & \\
$f_2(1270)$  & $0^+ (2^{++})$ & $1275.5 \pm 0.8$ & $2.6 \pm 0.5$ \\
$a_2(1320)$  & $1^- (2^{++})$ & $1318.3^{+0.5}_{-0.6}$ & $1.00 \pm 0.06$ \\
\hline
\end{tabular}
\end{center}
\end{table}

The invariant matrix elements describing the 
$e^+ e^- \to \gamma^* \to \gamma S$ 
and $e^+ e^- \to \gamma^* \to \gamma T$ annihilation processes  are given by 
\eq
M(e^+ e^- \to \gamma^* \to \gamma S) &=& - e^3 \, g_{S\gamma\gamma}
\bar v(p_2) \gamma_\mu u(p_1) \, \epsilon_\nu^{* \lambda}(q_1) \,
(g^{\mu\nu} qq_1 - q^\nu q_1^\mu) \, \frac{F_{\gamma^*\gamma S}(s)}{s} \,,
\nonumber\\
M(e^+ e^- \to \gamma^* \to \gamma T) &=& - e^3 \, g_{T\gamma\gamma}
\bar v(p_2) \gamma_\alpha u(p_1) \, \epsilon_\sigma^{* \lambda}(q_1)
\, \epsilon_{\mu\nu}^{* \lambda_T}(q_2) \,
\Big(g_{\rho}^\alpha q^{\mu} - g^{\mu\alpha} q_{\rho}\Big)
\, \Big(g^{\rho\sigma} q_1^\nu - g^{\sigma\nu} q_1^\rho\Big)
\, \frac{F_{\gamma^*\gamma T}(s)}{s} \,,
\en
where $\epsilon_\nu^{* \lambda}$ and 
$\epsilon_{\mu\nu}^{* \lambda_T}$ are the polarization vectors 
of final photon and tensor meson, respectively. 
The corresponding differential cross sections are derived as 
\eq
\frac{d\sigma}{dt}(e^+ e^- \to \gamma S) &=&
\frac{4 \pi \alpha \Gamma(S \to \gamma\gamma)}{M_S^3} \
\frac{|F_{\gamma^*\gamma S}(s)|^2}{s} \
\biggl[ 1 + \frac{2 t}{s} - \frac{2 M_S^2}{s}
+ \frac{2 t^2}{s^2} - \frac{2 M_S^2 t}{s^2}
+ \frac{M_S^4}{s^2} \biggr] \,, \nonumber\\
\frac{d\sigma}{dt}(e^+ e^- \to \gamma T) &=&
\frac{10 \pi \alpha \Gamma(T \to \gamma\gamma)}{3 M_T^3} \
\frac{|F_{\gamma^*\gamma T}(s)|^2 s}{M_T^4} \
\biggl[ 1 + \frac{2 t}{s}
- \frac{2 M_T^2}{s}
+ \frac{2 t^2}{s^2}
+ \frac{7 M_T^4}{s^2}
- \frac{14 t M_T^2}{s^2} \nonumber\\
&-& \frac{12 t^2 M_T^2}{s^3}
+ \frac{24 t M_T^4}{s^3}
- \frac{12 M_T^6}{s^3}
+ \frac{12 t^2 M_T^4}{s^4}
- \frac{12 t M_T^6}{s^4}
+ \frac{6 M_T^8}{s^4}
\biggr] \,.
\en
As dictated by the QCD compositeness, the $1/s^3$ scaling of the differential 
cross sections for the production of scalar and tensor mesons as 
quark-antiquark systems requires the following scaling of 
the $F_{\gamma^*\gamma S}(s)$ and 
$F_{\gamma^*\gamma T}(s)$ transition form factors at large $s$
\eq\label{scaling_FF} 
F_{\gamma^*\gamma S}(s) \sim \frac{1}{s}\,, \quad
F_{\gamma^*\gamma T}(s) \sim \frac{1}{s^2} \,. 
\en 
These results are hold when scalar and tensor mesons are 
systems of two tightly bound diquark in a possible CD tetraquark 
scenario, while for other tetraquarks we get 
extra falloffs $1/s$ and $1/s^2$ for $F_{\gamma^*\gamma S}(s)$ and 
$F_{\gamma^*\gamma T}(s)$, respectively.  
The integral cross sections are in agreement 
with Refs.~\cite{Achasov:2013daa,Achasov:2015pba} and are given by 
\eq
\sigma(e^+ e^- \to \gamma S) &=&
\frac{8 \pi \alpha \, \Gamma(S \to \gamma\gamma)}{3 M_S^3} \,
\Big|F_{\gamma^*\gamma S}(s)\Big|^2 \, \biggl(1 - \frac{M_S^2}{s}\biggr)^3
\,, \nonumber\\
\sigma(e^+ e^- \to \gamma T) &=&
\frac{20 \pi \alpha \, \Gamma(T \to \gamma\gamma)}{9 M_T^3} \,
\biggl|\frac{s}{M_T^2} \, F_{\gamma^*\gamma T}(s)\biggr|^2
\, \biggl(1 - \frac{M_T^2}{s}\biggr)^3 \,
\biggl[ 1 + \frac{3 M_T^2}{s} + \frac{6 M_T^4}{s^2} \biggr] \,.
\en
The total cross sections scale as $1/s^2$ when the corresponding 
form factors scale according to~(\ref{scaling_FF}). 
In particular, the asymptotic expressions for integral cross sections 
at leading order in $1/s$ expansion for scalar and tensor mesons read 
\eq
\sigma_{\rm Asymp}(e^+ e^- \to \gamma S) &=&
\frac{8 \pi \alpha \, \Gamma(S \to \gamma\gamma)}{3 M_{S}^3} \,
\frac{\Lambda_{S}^4}{s^2} \,, 
\label{sigma_asymptS}\\
\sigma_{\rm Asymp}(e^+ e^- \to \gamma T) &=&
\frac{20 \pi \alpha \, \Gamma(T \to \gamma\gamma)}{9 M_{T}^3} \,
\frac{\Lambda_{T}^4}{s^2} 
\label{sigma_asymptT}\,. 
\en 
First, we focus on extraction of transition form factors for 
scalar and tensor mesons. In case of scalar mesons, 
using production data on $f_0(980)$ and $a_0(980)$ we deduce the behavior 
of the $F_{\gamma^*\gamma f_0(980)}(s)$ and $F_{\gamma^*\gamma a_0(980)}(s)$ 
form factors depending on the variable $s$ (total energy squared). 
For two scenarios --- scalar mesons $f_0(980)$ and $a_0(980)$ 
are quark-antiquark systems and tetraquark as a system 
of two tightly bound color diquarks we use the double-pole expression 
for these form factors, which has 3 free parameters --- 
scale parameter $\Lambda$ 
and two dimensionless parameters $a$ and $b$, fixed from a fit to the data: 
\eq 
F_{\gamma^*\gamma S}(s) = \frac{1 + a \hat{s}} 
{1 - b \hat{s} + a \hat{s}^2}\,, \quad \hat{s} = \frac{s}{\Lambda^2} \,. 
\en 
The $F_{\gamma^*\gamma S}(s)$ form factor is normalized 
to 1 at $s=0$ and displays the $1/s$ scaling for large $s$, 
as required by the QCD compositeness, with 
\eq 
F_{\gamma^*\gamma S}(s) \sim \frac{\Lambda^2}{s} \,. 
\en  
Therefore, the value of the scale parameter $\Lambda$ must be fixed 
from the asymptotic behavior of the cross sections for $\gamma f_0$ and 
$\gamma a_0$ production. 
Using the available data from Ref.~\cite{Achasov:2013daa}, we fix the values 
of $\Lambda$ for $\gamma f_0$ and $\gamma a_0$ production as 
\eq 
\Lambda_{f_0} = 350 \pm 50 \ {\rm MeV}\,, \quad 
\Lambda_{a_0} = 250 \pm 50 \ {\rm MeV}\,. 
\en 
In this case the parameters $a$ and $b$ are fixed as 
\eq
a = 0.011^{+0.008}_{-0.005}\,, \quad  
b = 0.190^{+0.063}_{-0.052}
\en
for the case of the $f_0$ form factor, while we have
\eq
a = 0.003^{+0.003}_{-0.002}\,, \quad  
b = 0.090^{+0.042}_{-0.033}
\en
for the case of the $a_0$.
For the ratio of the $e^+ e^- \to \gamma f_0(980)$ and 
$e^+ e^- \to \gamma a_0(980)$ cross sections at large $s$ 
we get 
\eq 
\frac{\sigma(e^+ e^- \to \gamma f_0(980))} 
     {\sigma(e^+ e^- \to \gamma a_0(980))} = 
\biggl(\frac{\Lambda_{f_0}}{\Lambda_{a_0}}\biggr)^4 \, 
\biggl(\frac{M_{a_0}}{M_{f_0}}\biggr)^3 \, 
\frac{\Gamma(f_0 \to \gamma\gamma)}{\Gamma(a_0 \to \gamma\gamma)}  
\simeq  \biggl(\frac{\Lambda_{f_0}}{\Lambda_{a_0}}\biggr)^4 \,. 
\en 

\begin{figure}[htb]
\begin{center}
\epsfig{figure=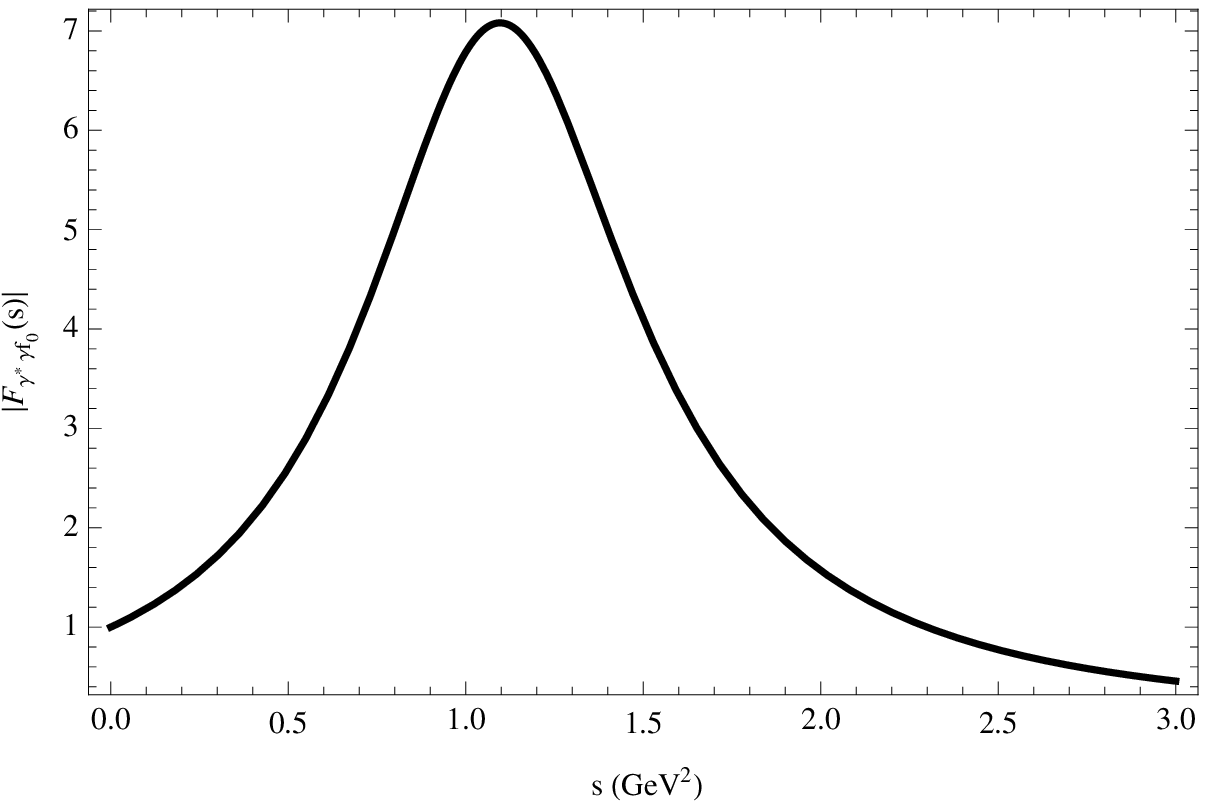,scale=.5}\hspace*{.5cm}
\epsfig{figure=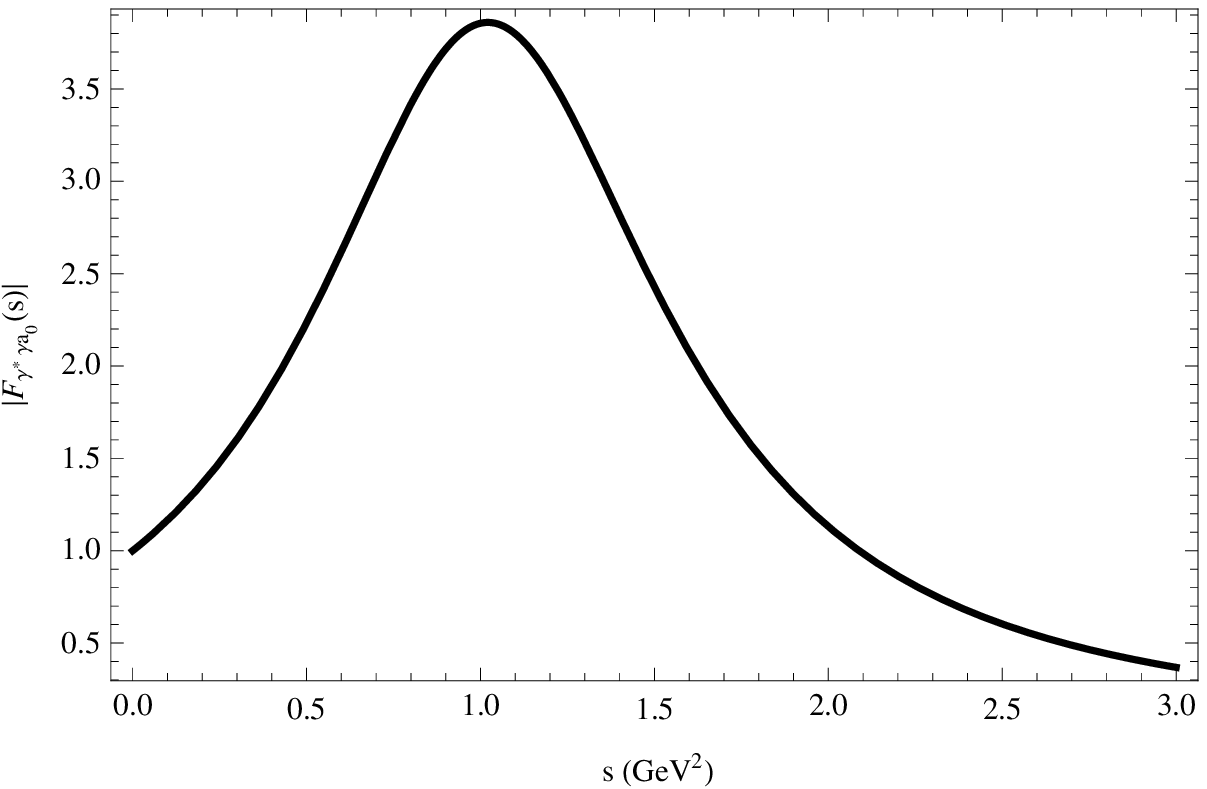,scale=.5}
\noindent
\caption{Form factors $F_{\gamma^*\gamma f_0(980)}(s)$ 
and $F_{\gamma^*\gamma a_0(980)}(s)$ 
at $\Lambda_{f_0} = 350$ MeV and $\Lambda_{a_0} = 250$ MeV.
\label{fig:Gf0a0gg}}
\end{center}
\begin{center}
\epsfig{figure=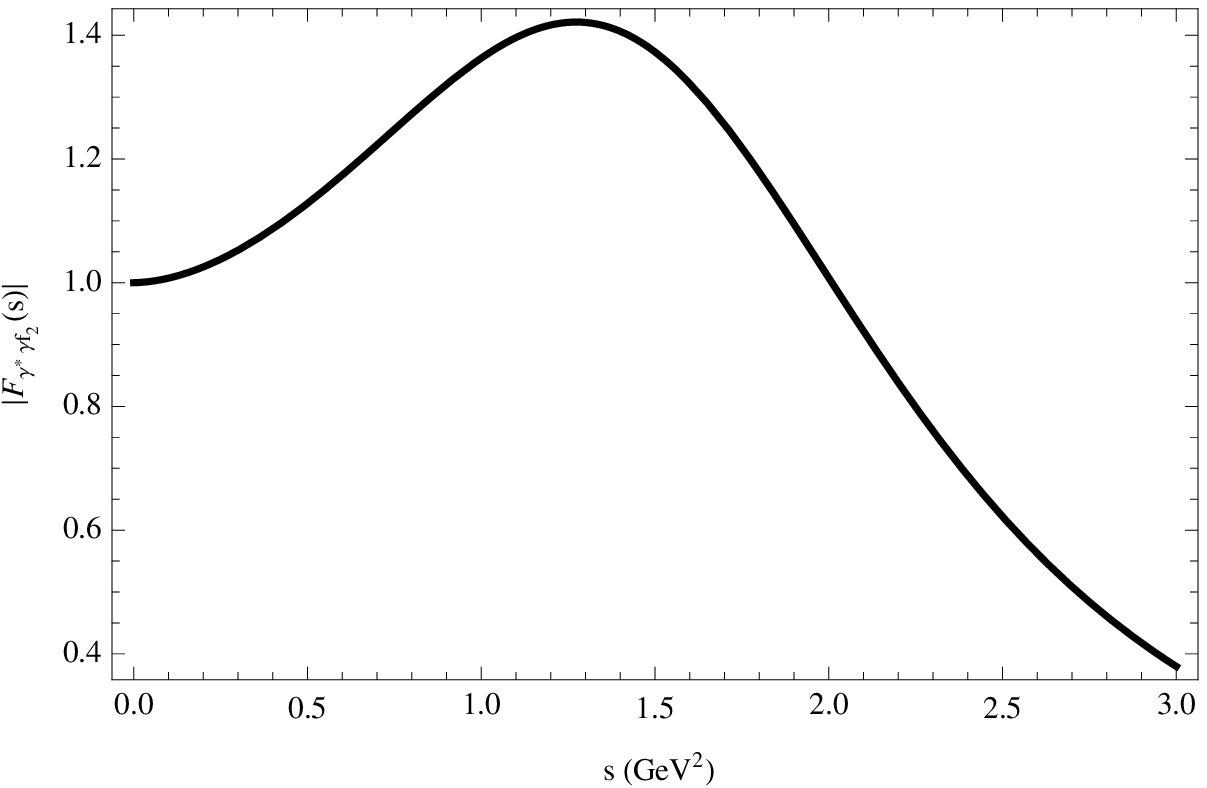,scale=.5}\hspace*{.5cm}
\epsfig{figure=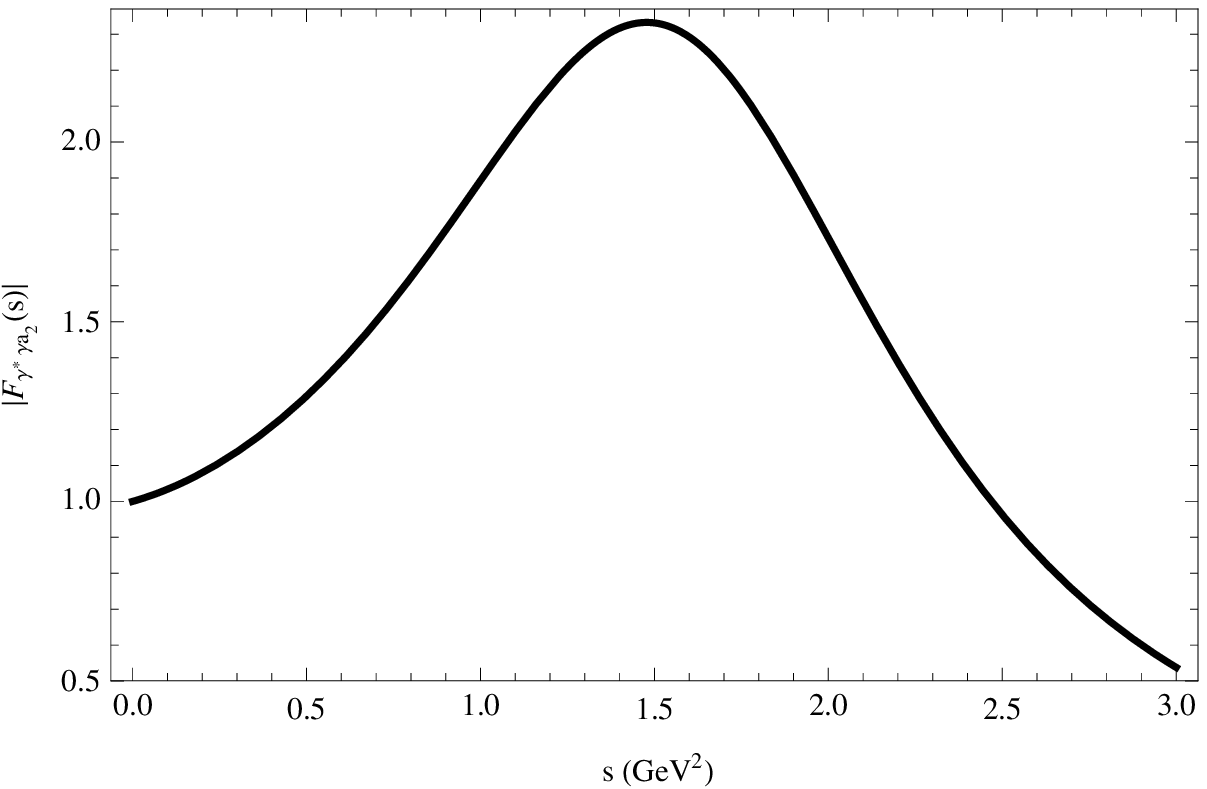,scale=.5}
\noindent
\caption{Form factors $F_{\gamma^*\gamma f_2(1270)}(s)$ 
and $F_{\gamma^*\gamma a_2(1320)}(s)$ 
at $\Lambda_{f_2} = 950$ MeV and $\Lambda_{a_2} = 1000$ MeV.
\label{fig:Gf2a2gg}}
\end{center}
\end{figure}

Plots of the form factors 
$F_{\gamma^*\gamma f_0(980)}(s)$ and 
$F_{\gamma^*\gamma a_0(980)}(s)$ are shown in Fig.~2. 
In case of tensor mesons we have only results of phenomenological 
consideration in Ref.~\cite{Achasov:2013daa}. 
We use the predictions of Ref.~\cite{Achasov:2013daa} for integral 
cross sections of production of tensor mesons to constraint the 
$\gamma^* \to \gamma T$ transition form factors. Taking into account 
the $1/s^2$ scaling of these form factors at large $s$ we use for them 
the following parametrization
\eq 
F_{\gamma^*\gamma T}(s) = \frac{1}{1 + \hat{s}} \, \frac{1 + a \hat{s}} 
{1 - b \hat{s} + a \hat{s}^2}\,, \quad \hat{s} = \frac{s}{\Lambda^2} \,. 
\en 
In case of the specific tensor meson states 
$f_2(1270)$ and $a_2(1320)$ the parameters $a$,$b$ and $\Lambda$ 
are fixed as
\eq
a = 0.241\,, \quad  
b = 0.771\,, \quad \Lambda = 950 \ {\rm MeV} 
\en
for the case of the $f_2(1270)$ form factor, while we have
\eq
a = 0.310\,, \quad  
b = 0.964\,, \quad \Lambda = 1000 \ {\rm MeV}  
\en
for the case of the $a_2(1320)$. The plots of the 
$\gamma^* \to \gamma f_2(1270)$ and $\gamma^* \to \gamma a_2(1320)$ transition 
form factors are shown in Fig.~3. 

In Fig.~4 we show our results for the integral cross sections 
$\sigma(e^+ e^- \to \gamma f_0(980))$ and 
$\sigma(e^+ e^- \to \gamma a_0(980))$ and compare them
with data points of the OLYA~\cite{Ivanov:1981wf} 
and DM2~\cite{Bisello:1988ez} collaborations 
extracted in Ref.~\cite{Achasov:2013daa}. Also we show the curves 
for asymptotical cross sections [see Eq.~(\ref{sigma_asymptS})]. 
One can see that since values $\sqrt{s} \simeq 2.4$ GeV our exact results 
for integral cross sections in case of the $f_0(980)$ and $a_0(980)$  
coincide with asymptotical ones.  

In Figs.~5 and 6 we present our results for other isoscalar 
[$f_0$--family: $f_0(1370)$, $f_0(1500)$, $f_0(1710)$] 
and isovector 
[$a_0$--family: $a_0(980)$, $a_0(1450)$, $a_0(1950)$] states. 
In case of $f_0$ family we present the results the ratio of the cross section 
and two-photon decay width $\Gamma(f_0 \to \gamma\gamma)$ 
because there are no data for the $\Gamma(f_0 \to \gamma\gamma)$. 
Note that 
the form factors for the $f_0$ and $a_0$ families are chosen as for 
the case of the $f_0(980)$ and $a_0(980)$, respectively, otherwise assuming
the same QCD compositeness for all the scalar state. 
One can see that perturbative regime there asymptotical cross sections 
coincide with exact calculation starts approximately from 
$\sqrt{s} \simeq  4$ GeV  
in case of the $f_0(1370)$, $f_0(1500)$, and $a_0(1450)$ states 
and from $\sqrt{s} \simeq  5$ GeV and $\sqrt{s} \simeq  5.5$ GeV in case 
of the $f_0(1710)$ and $a_0(1950)$ states, respectively. 

Finally, in Fig.~7 we present our predictions for the integral cross sections 
of tensor mesons. For a comparison we present 
results for asymptotic cross sections given by Eq.~(\ref{sigma_asymptS}). 
One can see that for tensor meson the perturbative regime starts 
from $\sqrt{s} \simeq 3.5$ GeV.  

In conclusion, we summarize the main results of the paper. 
Using quark counting rules for the differential cross section of 
the scalar and tensor mesons in single-photon annihilation processes 
we constrained the scaling of the $\gamma^* \to \gamma S$ and 
$\gamma^* \to \gamma T$ transition form factors. We showed 
that in case of quark-antiquark picture or tetraquark system of two tightly 
bound color diquarks they should scale as 
$1/s$ and $1/s^2$ for large $s$, respectively. 
In case of other tetraquark configurations and two-hadron molecule 
configurations according to counting of the constituents (this result  
is known from Refs.~\cite{Blitz:2015nra,Brodsky:2015wza,Brodsky:2016uln})      
they get extra $1/s$ falloff. 
Restricting to the first choice (quark-antiquark picture or 
tetraquark system of two tightly bound color diquarks) 
we extracted from data the behavior of these form factors with respect 
to the $s$ variable, using 
a double-pole formula with 3 free parameters. Using the obtained form factors 
for $f_0(980)$ and $a_0(980)$ we predict the integral cross sections 
for the members of their families --- 
isoscalar [$f_0(1370)$, $f_0(1500)$, $f_0(1710)$] and isovector 
[$a_0(1450)$, $a_0(1950)$] states. Here we assumed that the transition 
form factors do not depend on the mass of the corresponding scalar meson.  
Similar analysis done for the tensor states $f_2(1270)$ and $a_2(1320)$. 
We think that our predictions for the transition form factors and integral 
cross sections of scalar and tensor mesons 
will be useful for future experiments at electron-positron colliders. 

\clearpage 

\begin{figure}[htb]
\epsfig{figure=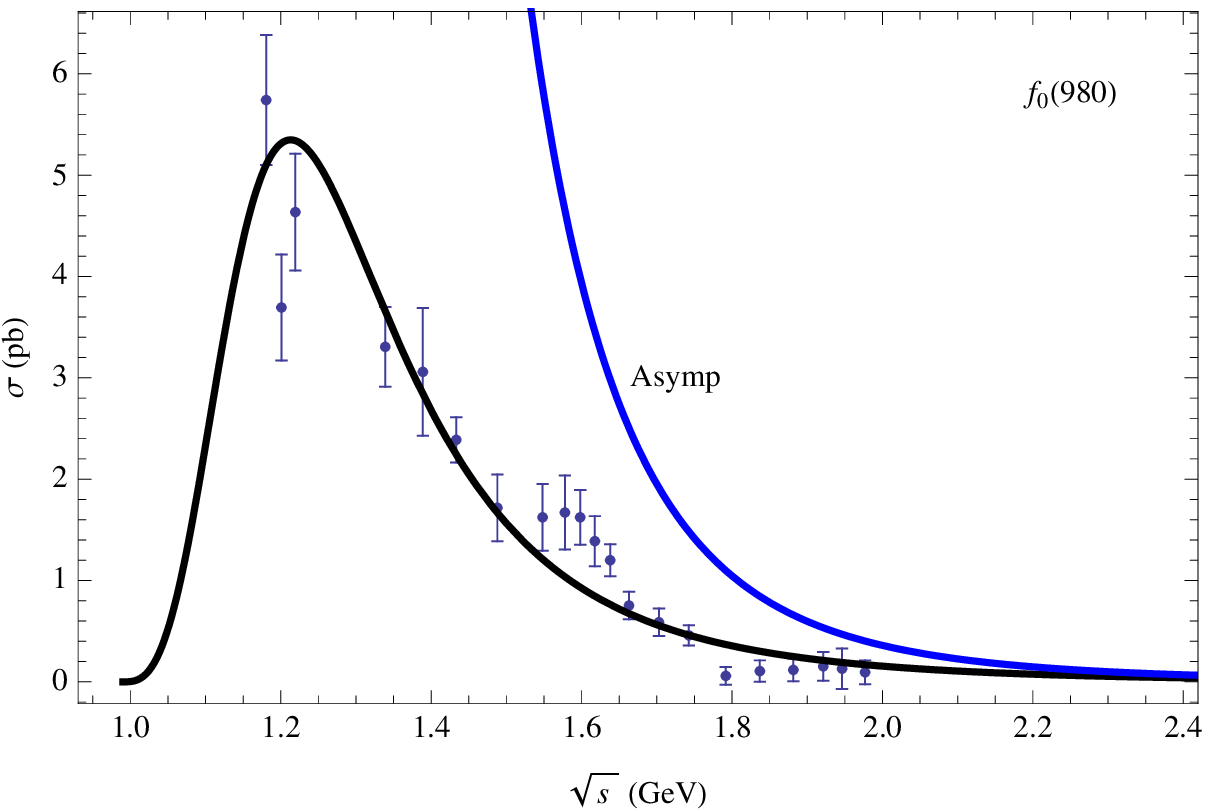,scale=.5}\hspace*{.5cm}
\epsfig{figure=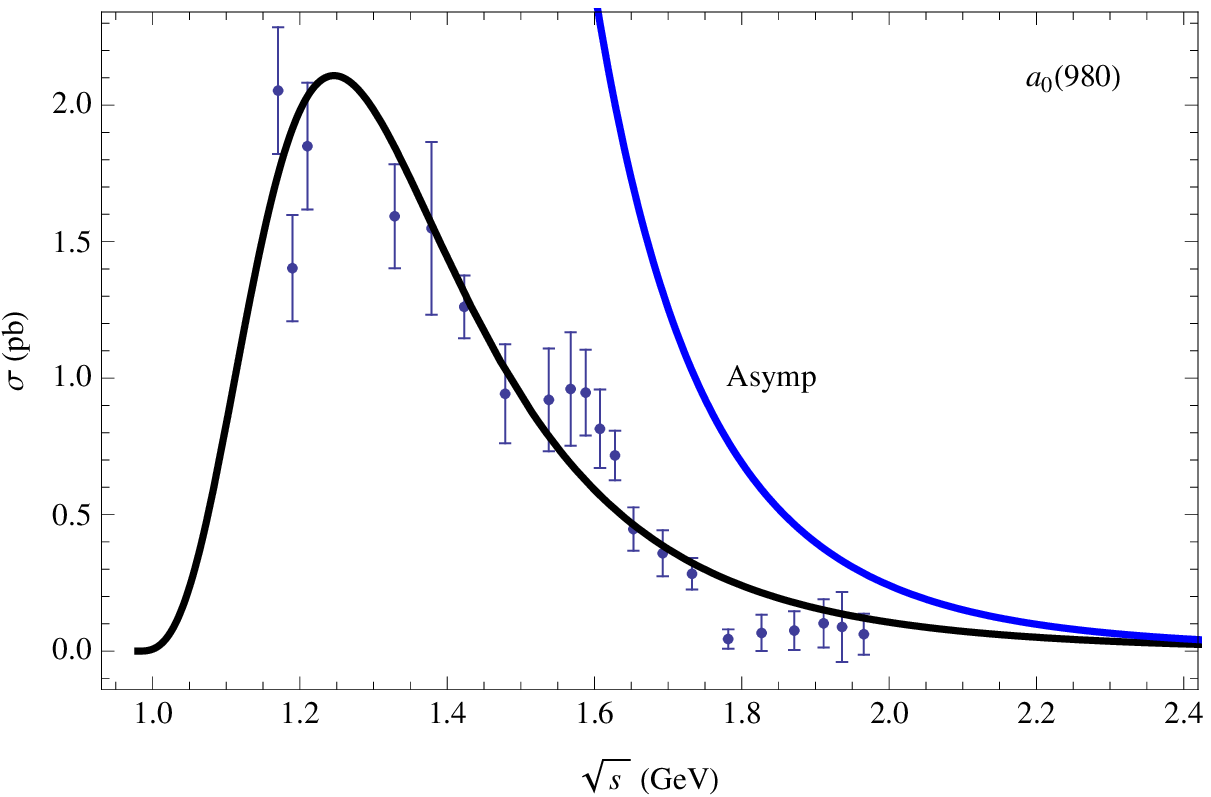,scale=.5}
\noindent
\caption{$\sigma(e^+ e^- \to \gamma^* \to \gamma f_0(980))$ and 
$\sigma(e^+ e^- \to \gamma^* \to \gamma a_0(980))$ 
at $\Lambda_{f_0} = 350$ MeV and $\Lambda_{a_0} = 250$ MeV.
Data are taken from Ref.~\cite{Achasov:2013daa}. 
\label{fig:f0a0}}
\vspace*{.5cm}
\epsfig{figure=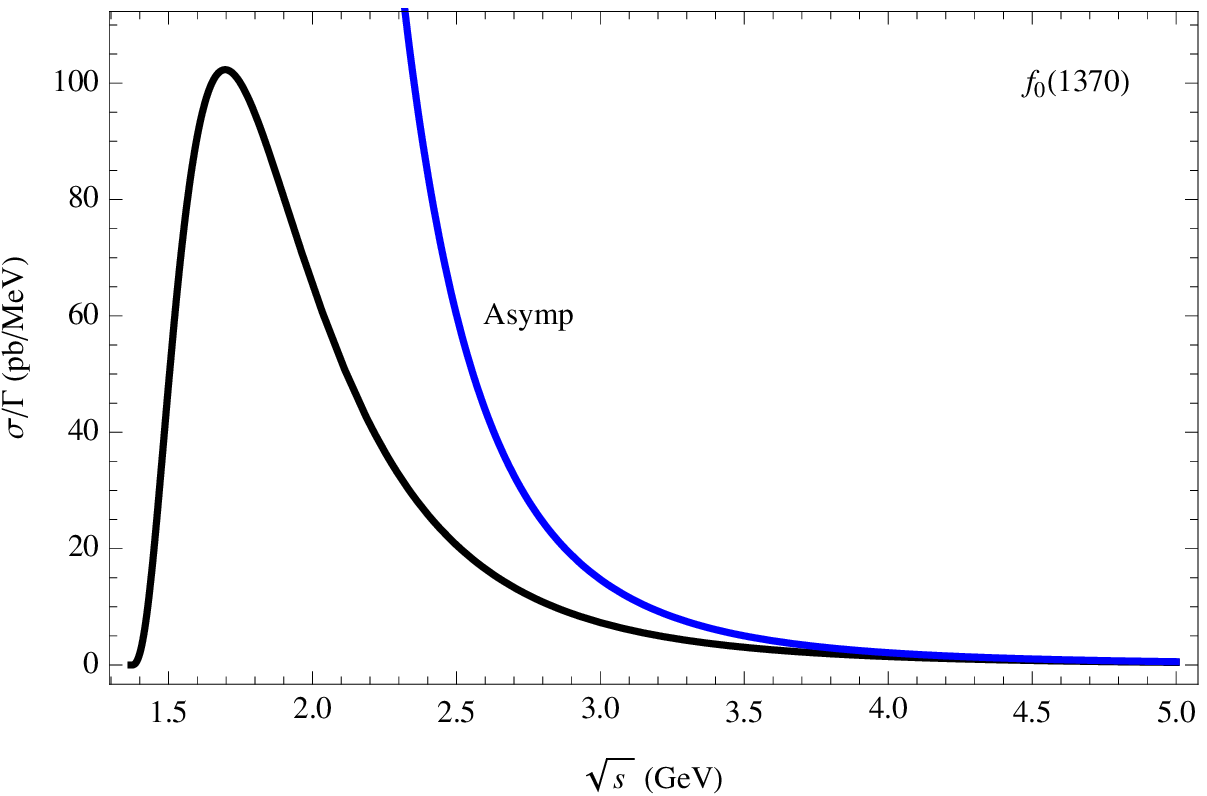,scale=.45}\, 
\epsfig{figure=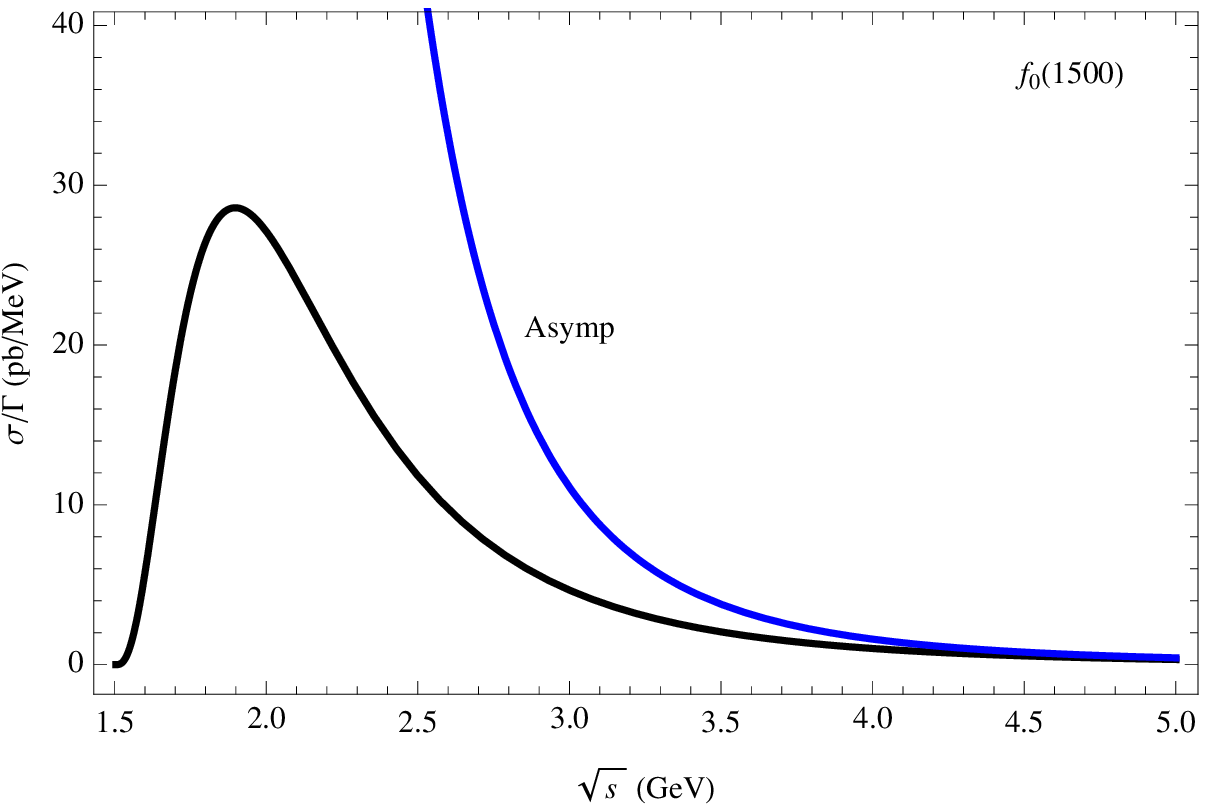,scale=.45}\, 
\epsfig{figure=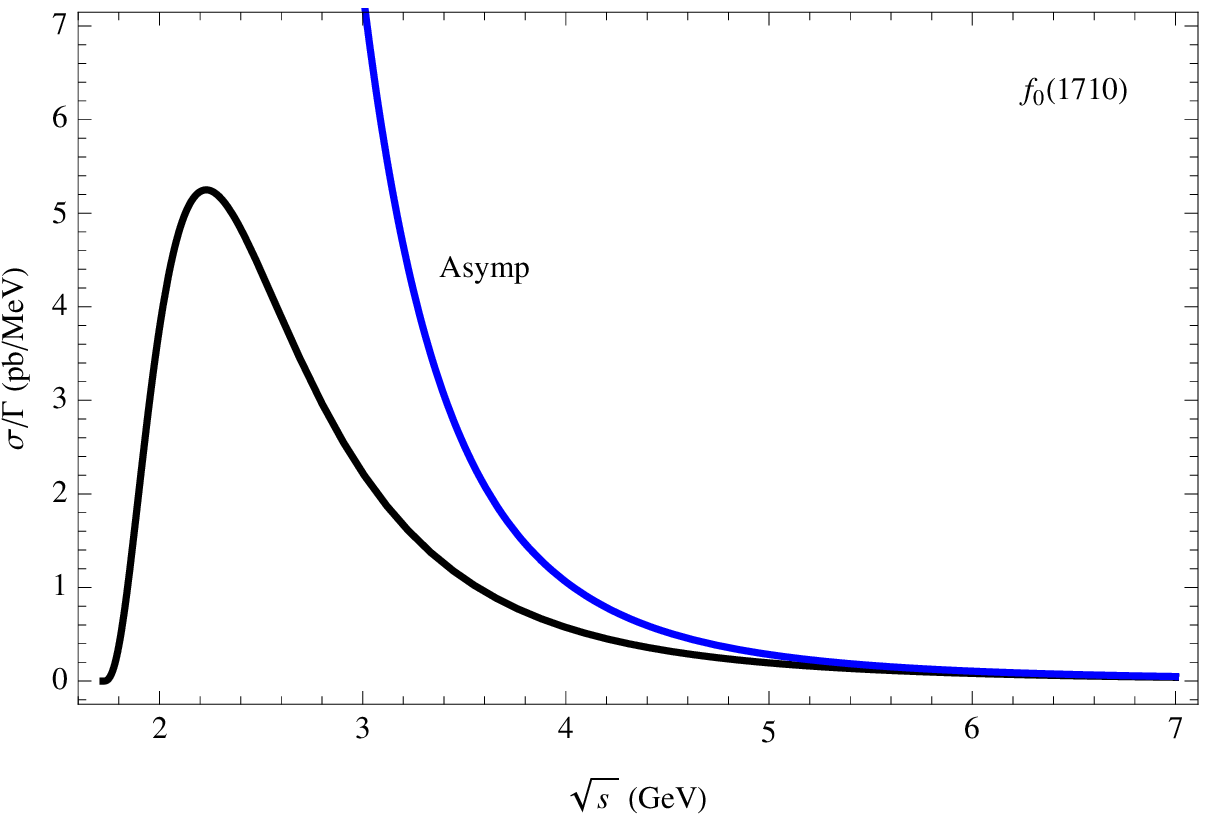,scale=.45}
\noindent
\caption{
The integral cross section divided by two-photon decay width 
$\sigma(e^+ e^- \to \gamma^* \to \gamma f_0)/
\Gamma(f_0 \to \gamma\gamma)$ 
for $f_0 = f_0(1370)$, $f_0(1500)$, and $f_0(1710)$   
with $\Lambda_{f_0} = 350$ MeV. 
\label{fig:allf0sigma}}
\vspace*{.5cm}
\epsfig{figure=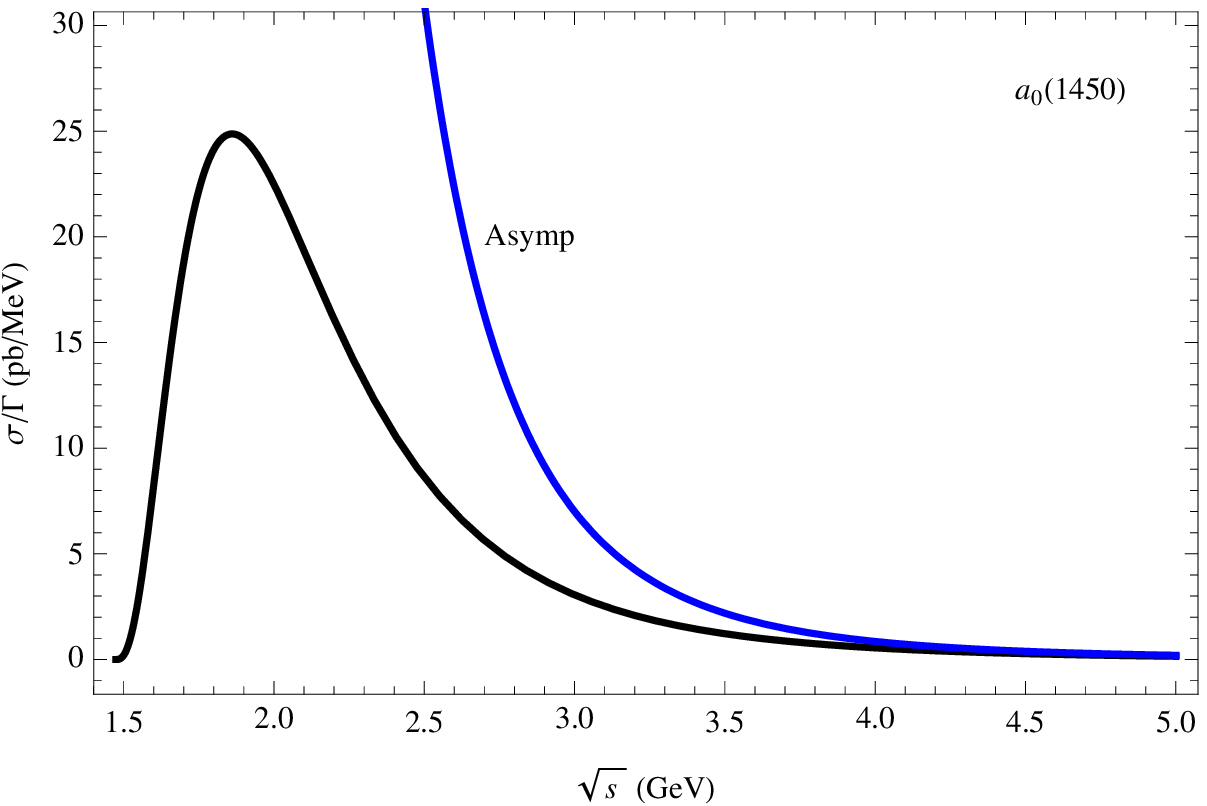,scale=.5}\, 
\epsfig{figure=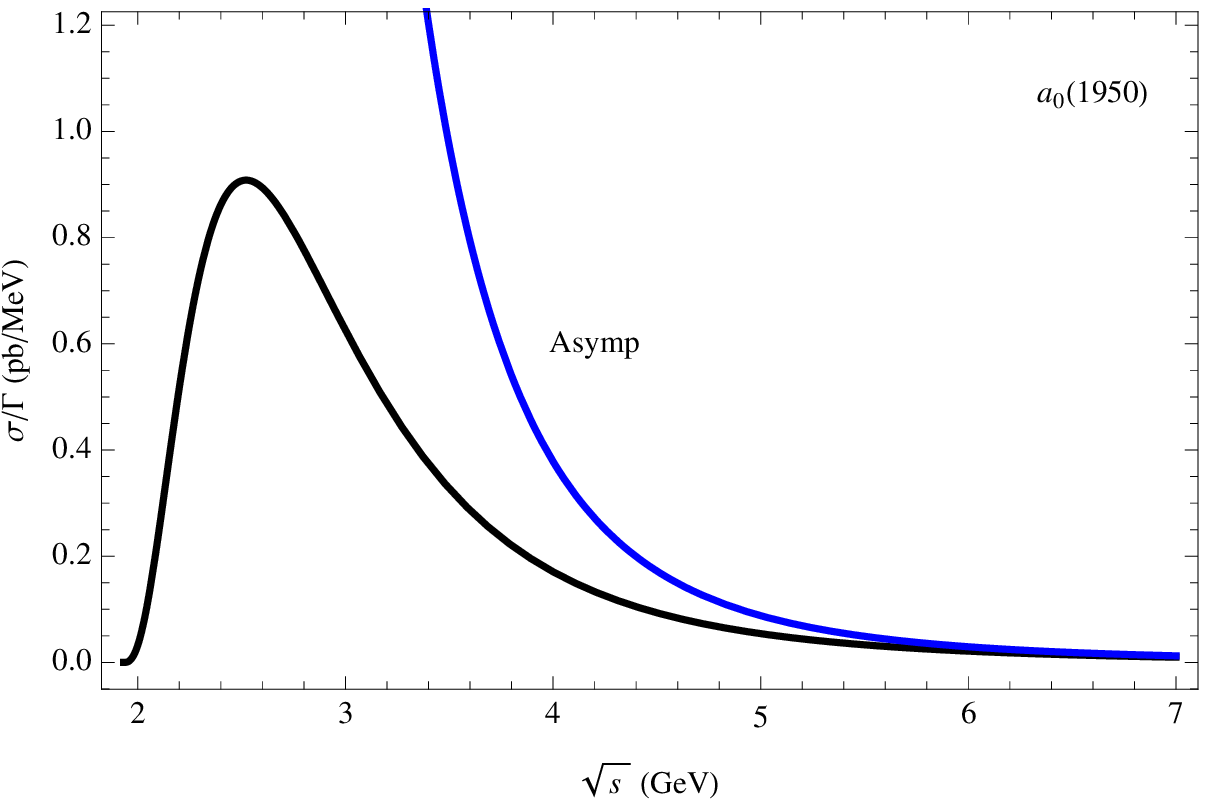,scale=.5}
\noindent
\caption{
The integral cross section divided by two-photon decay width 
$\sigma(e^+ e^- \to \gamma^* \to \gamma a_0)/
\Gamma(a_0 \to \gamma\gamma)$ 
for $a_0 = a_0(1450)$ and $a_0(1950)$ 
with $\Lambda_{a_0} = 250$ MeV. 
\label{fig:alla0sigma}}
\vspace*{.5cm}
\epsfig{figure=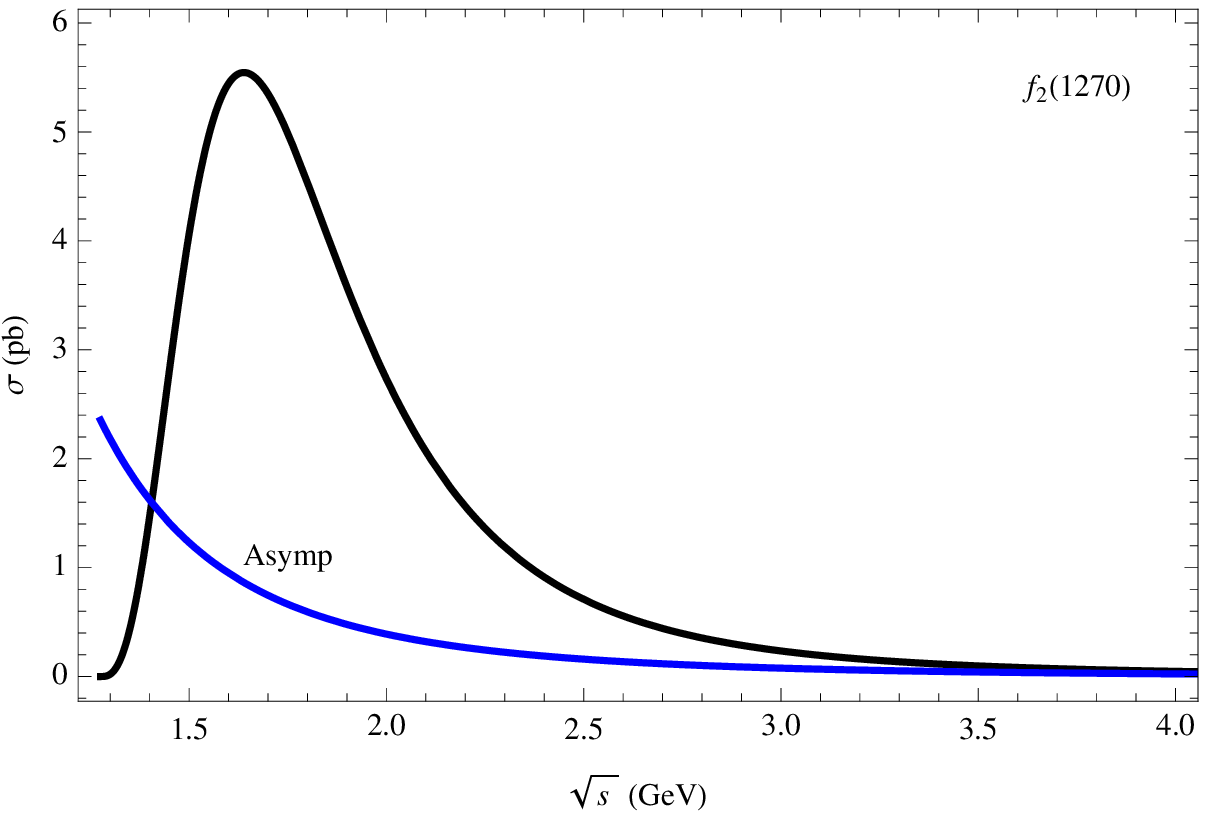,scale=.5}\hspace*{.5cm}
\epsfig{figure=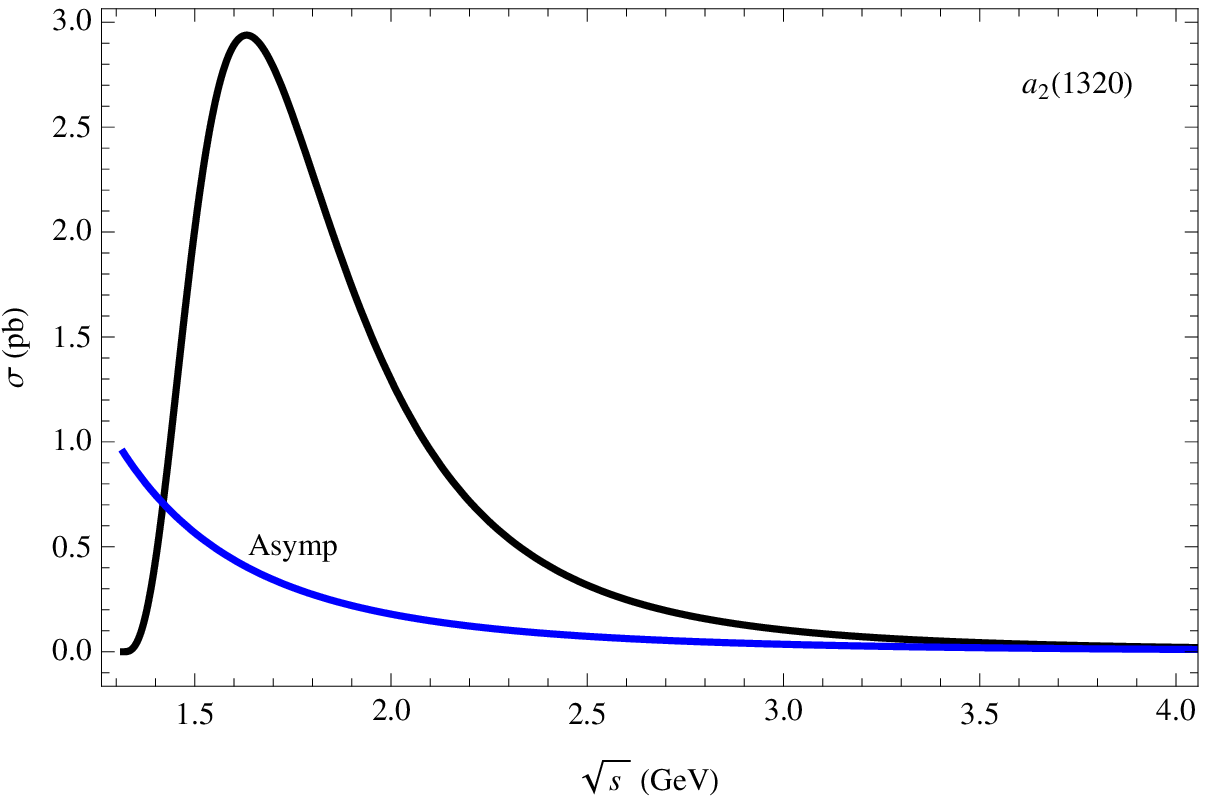,scale=.5}
\noindent
\caption{$\sigma(e^+ e^- \to \gamma^* \to \gamma f_2(1270))$ and 
$\sigma(e^+ e^- \to \gamma^* \to \gamma a_2(1320))$ 
at $\Lambda_{f_2} = 950$ MeV and $\Lambda_{a_2} = 1000$ MeV.
\label{fig:f2a2sigma}}
\end{figure}

\clearpage 

\begin{acknowledgments}

This work was funded
by the German Bundesministerium f\"ur Bildung und Forschung (BMBF)
under Project 05P2015 - ALICE at High Rate (BMBF-FSP 202):
``Jet- and fragmentation processes at ALICE and the parton structure
of nuclei and structure of heavy hadrons'';
by CONICYT (Chile) PIA/Basal FB0821,
by CONICYT (Chile) Research Projects No. 80140097,
and under Grants No. 7912010025, 1140390;
by Tomsk State University Competitiveness
Improvement Program, by the Russian Federation program ``Nauka''
(Contract No. 0.1764.GZB.2017), by Tomsk Polytechnic University 
Competitiveness Enhancement Program grant (Grant No. VIU-FTI-72/2017).

\end{acknowledgments}

\end{document}